\documentclass[pra,aps]{revtex4}
\usepackage{amsmath}
\usepackage{graphicx}

\begin{document}

\title{Computer simulation of two-mode nonlinear quantum scissors}

\author{T. Dung Nguyen }
\author{W. Leo\'nski}
\email{wleonski@proton.if.uz.zgora.pl}
\author{V. Cao Long}
\affiliation{Quantum Optics and Engineering Division, Institute of Physics, University of Zielona G\'ora, A. Szafrana 4a, 65-516 Zielona G\'ora, Poland}

\date{\today}

\begin{abstract}
We present a simulation method allowing for modeling of quantum dynamics of nonlinear quantum scissors' (NQS) systems. We concentrate on the two-mode model involving two mutually interacting nonlinear quantum oscillators (Kerr nonlinear coupler) excited by a series of ultra-short external coherent pulses. We show that despite the simplicity of the method one can obtain non-trivial results. In particular, we discuss and compare two cases of kicked nonlinear coupler, showing that the quantum evolution of the system remains closed within a two-qubit Hilbert space and can lead to maximally entangled states generation.  
\end{abstract}
\keywords{quantum state engineering, Kerr coupler, nonlinear quantum scissors, qubit}
\maketitle

\pagenumbering{arabic}

\section{Introduction}
Quantum states engineering is one of the most interesting fields in quantum optics. It seems to be very promising, especially from the point of view of quantum information theory. Some optical systems, including nonlinear media allow for generation of states that are defined in finite-dimensional Hilbert space, and they are referred to as \textit{nonlinear quantum scissors} (NQS) -- for comprehensive discussion of such systems see for instance \cite{LK11} \textit{and the references quoted therein}). Since such models involve nonlinear subsystems, finding analytical solutions describing their dynamics can be cumbersome. Therefore, for such cases numerical methods could become only one possible way for solving given problem. In this paper we shall show how to model quantum dynamics of nonlinear system. In particular, using kicked nonlinear quantum coupler model we will show that the system's dynamics can be closed within a finite set of quantum states. In consequence,  our system can be treated as NQS.

\section{The model and simulation method}
In this paper we shall concentrate on the two-mode model comprising two mutually interacting quantum nonlinear oscillators labeled by $a$ and $b$. They are described by the following Hamiltonian expressed in terms of boson creation and annihilation operators $\hat{a}^\dagger$ ($\hat{b}^\dagger$) and $\hat{a}$ ($\hat{b}$), respectively:
\begin{equation}
\hat{H}_{NL}\ =\ \frac{\chi_a}{2}(\hat{a}^\dagger)^2\hat{a}^2+\frac{\chi_b}{2}(\hat{b}^\dagger)^2\hat{b}^2+
\epsilon\hat{a}^{\dagger}\hat{b}+\epsilon^*\hat{a}\hat{b}^{\dagger}+
\chi_{ab}\hat{a}^{\dagger}\hat{a}\hat{b}^{\dagger}\hat{b},
\label{eq1}
\end{equation}
where $\chi_a$ and $\chi_b$ are nonlinearity coefficients corresponding to two Kerr-like oscillators, $\epsilon$ describes the strength of their mutual linear interaction whereas $\chi_{ab}$ is so-called  \textit{cross-coupling term}. The latter is sometimes omitted by authors and the discussion of this problem can be found for example in \cite{KP97}. The model presented here is an extension of that discussed in \cite{ML06} and thanks to presence of the cross-coupling  term seems to be more realistic. Moreover, it differs from that considered previously in \cite{LKM10} in fact that we assume here linear interaction between two oscillators whereas for the system discussed there nonlinear coupling was assumed. 

The system is externally excited in one mode and this excitation is in the form of series of ultra-short coherent pulses. Such interaction between the coupler and external field can be modeled with use of \textit{Dirac-delta} functions. In consequence, the Hamiltonian corresponding to this interaction can be written as:
\begin{equation}\label{eq2}
\hat{H}_{K}=(\alpha\hat{a}^{\dag}+\alpha^*\hat{a})\,\sum\limits_{k=0}^{\infty}\delta(t-kT).
\end{equation}
The parameter $\alpha$ appearing here describes the strength of the external field - nonlinear system interaction, $k$ enumerates external pulses, whereas $T$ is a time between two subsequent pulses.

Since we neglect here all dissipation processes, time-evolution of the system can be described by wave-functions. It can be expressed with use of $n$-photon states. Our system is two-mode model and hence, can be defined as
\begin{equation}
|\Psi\rangle = \sum_{m,n=0}^{\infty} c_{m,n}|m\rangle_a\otimes|n\rangle_b,
\label{eq3}
\end{equation}
where $c_{m,n}$ are complex probability amplitudes, $|m\rangle_a$ and $|n\rangle_b$ are $n$-photon Fock states corresponding to the modes $a$ and $b$, respectively, and symbol $\otimes$ denotes Kronecker product. Due to the fact that $n$-photon states have discrete representation, they can be easily applied in numerical calculations. In fact, this wave function can be represented by a $m\times n$ element column vector containing complex probability amplitudes. It can be written as:
\begin{equation}
|\psi\rangle\,=\,\left[
\begin{array}{c}
c_{0,0}\\
c_{0,1}\\
c_{0,2}\\
\vdots\\
c_{0,n}\\
c_{1,0}\\
c_{1,1}\\
c_{1,2}\\
\vdots\\
c_{m,n},
\end{array}
\right]\label{eq4}
\end{equation}
where the normalization condition 
\begin{equation}
\sum\limits_{k=0}^{m}\sum\limits_{l=0}^{n}\,|c_{k,l}|^2=\,1
\end{equation}
should be fulfilled

For non-dissipative models time-evolution of the system is described by unitary evolution operators defined on the basis of the system's Hamiltonian. Since the interaction with external field is represented by ultra-short pulses, the whole evolution can be divided onto two subsequent, completely different in their nature stages. First of them corresponds to the period of time between two subsequent pulses. During this time the energy of the coupler is conserved and its evolution is determined by the following unitary operator
\begin{equation}
\hat{U}_{NL}=e^{-i\hat{H}_{NL}T}.
\label{eq5}
\end{equation}
The second stage corresponds to the interaction with a single pulse during infinitesimal short time and the operator responsible for this interaction can be written as
 \begin{equation}
\hat{U}_{K}=e^{-i(\alpha\hat{a}^{\dag}+\alpha^*\hat{a})}.
\label{eq6}
\end{equation}
In consequence, transformation of the wave-function from that corresponding to the moment just after $j$-th pulse to that after $j+1$ one, is a result of action of these two operators
\begin{equation}
|\psi_{j+1}\rangle\,=\,\hat{U}_K\,\hat{U}_{NL}\,|\psi_j\rangle.
\label{eq7}
\end{equation}
Thus, we can obtain the wave-function for the moment of time just after $k$-th pulse by multiple ($k$-times) action of the operator $\hat{U}=\hat{U}_K\hat{U}_{NL}$ on initial state of the system $|\psi_0\rangle$
\begin{equation}
|\psi_{k}\rangle\,=\,(\hat{U})^k\,|\psi_0\rangle\,=\,(\hat{U}_K\hat{U}_{NL})^k\,|\psi_0\rangle.
\label{eq8}
\end{equation}

Since we are using $n$-photon basis, we can define the creation (annihilation) operators as square matrices which in fact are sparse. For example, if we assume that the dimensions of Hilbert subspaces, corresponding to each of nonlinear oscillators are $n$ and $m$, respecitvely, the operator $\hat{a}$ can be written as
\begin{equation}
\hat{a}\,=\,\left[
\begin{array}{cccccc}
0 & 1 & 0 & \hdots & 0 & 0 \\
0& 0 &\sqrt{2} & \hdots& 0 & 0\\
0& 0 & 0 & \hdots& 0 & 0\\
\vdots & \vdots & \vdots & \ddots & \vdots & \vdots \\
0& 0 & 0 & \hdots & 0 & \sqrt{n-1}\\
0 & 0 & 0 & \hdots & 0 & 0
\end{array}
\right]\,\otimes\,\check{I}_m,\label{eq9}
\end{equation}
whereas its counterpart for the mode $b$ will be defined by
\begin{equation}
\hat{b}\,=\,\check{I}_n\,\otimes\,\left[
\begin{array}{cccccc}
0 & 1 & 0 & \hdots & 0 & 0 \\
0& 0 &\sqrt{2} & \hdots& 0 & 0\\
0& 0 & 0 & \hdots& 0 & 0\\
\vdots & \vdots & \vdots & \ddots & \vdots & \vdots \\
0& 0 & 0 & \hdots & 0 & \sqrt{m-1}\\
0 & 0 & 0 & \hdots & 0 & 0
\end{array}\right],\label{eq10}
\end{equation}
where $\check{I}_{n(m)}$ represents unity matrix which dimension is equal to $n$ ($m$), whereas symbol $\otimes$ denotes Kronecker product (the same as in (\ref{eq3})). Analogously, creation operators for the modes $a$ and $b$ can be expressed as
\begin{equation}
\hat{a}^\dagger\,=\,\left[
\begin{array}{cccccc}
0 & 0 & 0 & \hdots & 0 & 0 \\
1 & 0 & 0 & \hdots& 0 & 0\\
0 & \sqrt{2} & 0 &\hdots& 0 & 0\\
\vdots & \vdots & \vdots & \ddots & \vdots & \vdots \\
0& 0 & 0 & \hdots & 0 & 0\\
0 & 0 & 0 & \hdots & \sqrt{n-1} & 0
\end{array}
\right]\,\otimes\,\check{I}_m,\label{eq11}
\end{equation}
and
\begin{equation}
\hat{b}^\dagger\,=\,\check{I}_n\,\otimes\,\left[
\begin{array}{cccccc}
0 & 0 & 0 & \hdots & 0 & 0 \\
1 & 0 & 0 & \hdots& 0 & 0\\
0 & \sqrt{2} & 0 &\hdots& 0 & 0\\
\vdots & \vdots & \vdots & \ddots & \vdots & \vdots \\
0& 0 & 0 & \hdots & 0 & 0\\
0 & 0 & 0 & \hdots & \sqrt{n-1} & 0
\end{array}
\right],\label{eq12}
\end{equation}
respectively. Using such defined operators we can construct the Hamiltonians $\hat{H}_{NL}$ $\hat{H}_K$ and hence, unitary evolution operators $\hat{U}_{NL}$, $\hat{U}_K$ and $\hat{U}$ in matrix form.
What is important, the Hamiltonian $\hat{H}_{NL}$ is diagonal in $n$-photon basis. Therefore, the problem of derivation of the matrix corresponding to $\hat{U}_{NL}$ becomes straightforward. Moreover, as a result we get diagonal matrix again and hence, the problem of effectiveness of multiplication the wave-function (represented by a vector) by such matrix simplifies considerably. Unfortunately,  the Hamiltonian $\hat{H}_K$ is not diagonal. In consequence, to find the evolution operator $\hat{U}_K$ we need application of methods usually used for calculation of matrix exponentials. Nevertheless, one should keep in mind that the matrices corresponding to annihilation and creation operators are sparse. It is possible to use appropriate procedures defined for such sort of matrices, and their application improves the effectiveness of calculations. Fortunately, standard numerical procedures for the both: calculation of matrix exponentials and manipulation with sparse matrices are already implemented in various standard packages or libraries. For instance,  \textit{Matlab} computing environment and language \cite{Matlab} (or its free clone \textit{Octave} \cite{Octave}) seems to be suitable for our purposes due to their simplicity and ease of use. These software packages can be applied even by computer users who are not very experienced in numerical calculations.

\section{Results}
For our purposes we have assumed that the field was initially in the vacuum states for the both modes, \textit{i.e.} neither we have photons in the mode $a$ nor in $b$:
\begin{equation}
\label{eq13}
|\psi_0\rangle\,=\,|0\rangle_a\otimes|0\rangle_b
\end{equation}
Obviously, it is possible to assume that the system was initially in other quantum state. For instance, very often considered quantum-optical systems are assumed to be initially in coherent state  $|\alpha\rangle$ \cite{G63}. Such state can be expressed in $n$-photon basis as
\begin{equation}
\label{eq14}
|\alpha\rangle\,=\,\exp (-|\alpha|^2/2)\,\sum\limits_{n=0}^{\infty}
\,\dfrac{\alpha^n}{\sqrt{n!}}\,|n\rangle ,
\end{equation}
where $\alpha$ appearing in (\ref{eq14}) denotes there a complex variable, and is related to the mean number of photons by the following relation $\langle\hat{a}^\dagger\hat{a}\rangle = |\alpha |^2$.
Nevertheless, for our considerations we shall consider weak field limits and assume that initially we have practically no photons in the system. For such cases the condition $|\alpha |^2\ll 1 $ is fulfilled and hence, the initial field can be approximated by the product of vacuum states defined for each mode. That means that only one probability amplitude appearing in the vector (\ref{eq4}) describing initial state will be different from zero, \textit{i.e.} $c_{0,0}=1$. 

Applying multiple action of the unitary evolution operator $\hat{U}$ on the initial state $|\psi_0\rangle$  we obtain vectors comprising probability amplitudes for all considered basis states $|n\rangle_a\otimes |m\rangle_b$ corresponding to the moments of time just after subsequent external pulses. For our purposes we define the matrices for wave-function and operators where $10$ $n$-photon states are involved for each mode. Moreover, we assume that $\chi_{ab}=1$ and $\chi_{a}=\chi_{b}=1$. As it was mentioned earlier, we deal here with the cases of weak external excitation and internal coupling ($\alpha =1/25$ and $\epsilon =1/100$) -- all energies and strengths of interactions are expressed here in units of nonlinearity constants.

\begin{figure}[h]
\centerline{\includegraphics[width= \textwidth]{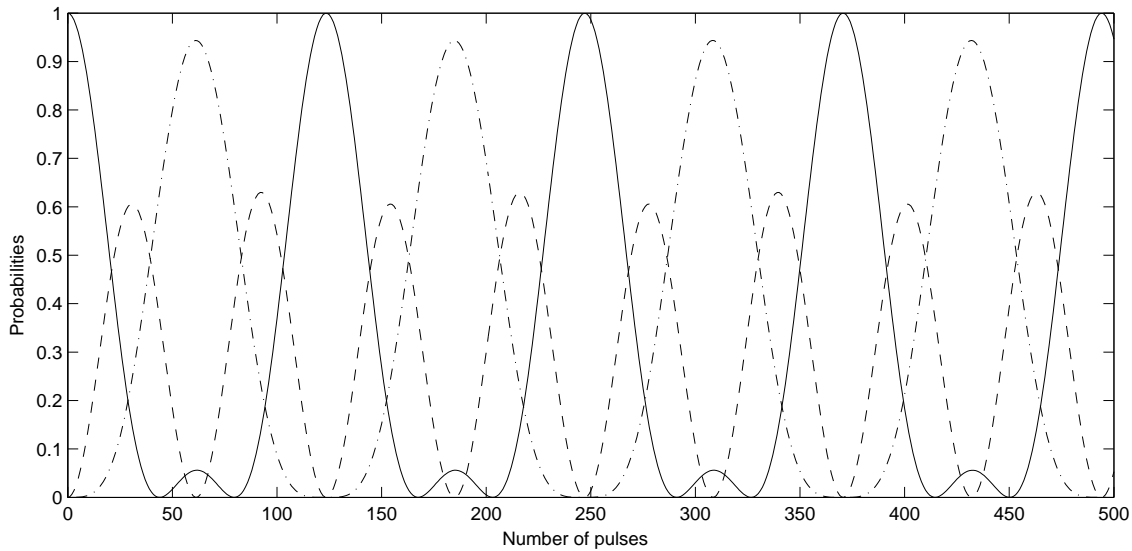}}
\caption {Probabilities for the states: $|0\rangle_a\otimes |0\rangle_b$ (solid line), $|0\rangle_a\otimes |1\rangle_b$ (dashed-dotted line) and $|1\rangle_a\otimes |0\rangle_b$ (dashed line). The parameters are: $T=\pi$, $\alpha =1/25$, $\epsilon =1/100$ and $\chi_a=\chi_b=\chi_{ab}=1$.}
\end{figure}

We start our considerations from the case when cross-coupling is taken into account. From Fig.1 we see that
only few states $|n\rangle_a\otimes |m\rangle_b$ are involved in the system's evolution and the system behaves as NQS \cite{LK11} (examples of such models were discussed for instance in \cite{LT94,GSK09,AGK11,GSK12}). For the models discussed here it is an effect of resonant coupling by  the zero-frequency component of external excitation between some eigenstates generated by the Hamiltonian $\hat{H}_{NL}$. We here assume that  cross-coupling is present and hence, only three states are involved in the evolution. They are: $|0\rangle_a\otimes |0\rangle_b$, $|0\rangle_a\otimes |1\rangle_b$ and $|1\rangle_a\otimes |0\rangle_b$. All these states correspond to the same eigenenergy equal to zero. Fig.1 shows the stime-evolution for the probabilities corresponding to these states. In fact, the sum $|c_{0,0}|^2+|c_{0,1}|^2+|c_{1,0}|^2\simeq 1$ and hence, the probabilities corresponding to other states can be neglected within our approximation. We see from Fig.2 that deviation of this sum from the unity oscillates and the amplitude of these oscillations is  $\sim 10^{-4}$. In consequence,  the truncation of the wave function involving only three states can be observed with high accuracy. From the point of view of quantum information theory, one can say that our system behaves as \textit{qubit-qubit} one $(0,1)_a$, $(0,1)_b$, because for two modes we have two possibilities: vacuum or one-photon state.

\begin{figure}[h]
\centerline{\includegraphics[width=0.7 \textwidth]{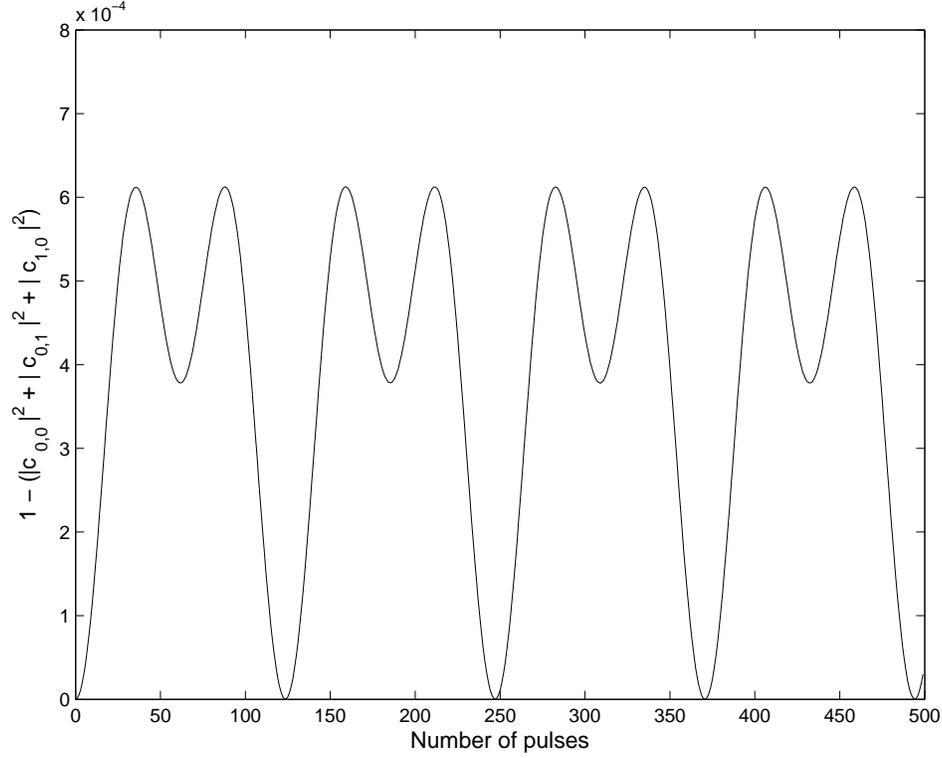}}
\caption{Deviation from the unity of the sum of probabilities shown in  Fig.1.}
\end{figure}

From Fig.1 one can see that for some moments of time the probabilities for the states $|0\rangle_a\otimes |0\rangle_b$ and $|1\rangle_a\otimes |0\rangle_b$ are simultaneously equal to $.5$. As a result, for such cases we get separable states  $1/\sqrt{2}(|0\rangle_a\otimes |0\rangle_b+ |1\rangle_a\otimes |0\rangle_b) =1/\sqrt{2}(|0\rangle_a+ |1\rangle_a ) \otimes |0\rangle_b$. What is more interesting, there are other moments of time for which  the both probabilities corresponding to the states $|0\rangle_a\otimes |1\rangle_b$ and $|1\rangle_a\otimes |0\rangle_b$ are equal to $.5$, as well. This situation corresponds to the generation of maximally entangled states -- \textit{Bell states.} For our model they are $|B_1\rangle=\frac{1}{\sqrt{2}}(|0\rangle_a|1\rangle_b+i|1\rangle_a|0\rangle_b)$ and $|B_2\rangle=\frac{1}{\sqrt{2}}(|1\rangle_a|0\rangle_b-i|0\rangle_a|1\rangle_b)$.

\begin{figure}[h]
\centerline{\includegraphics[width=0.65 \textwidth]{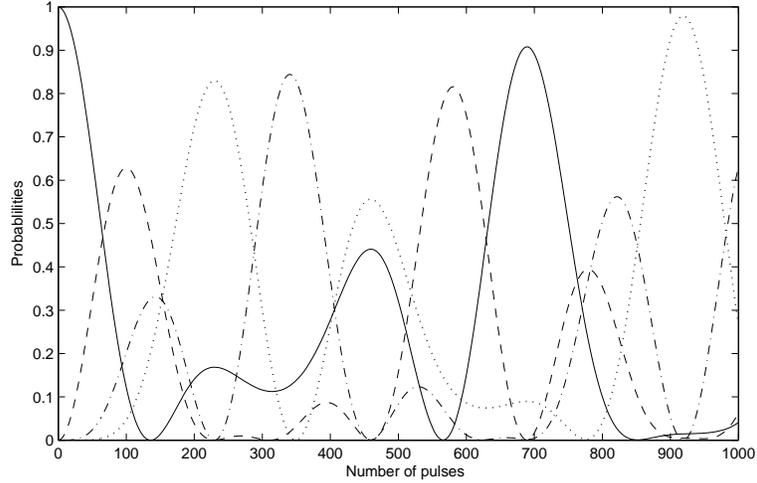}}
\caption{Probabilities for the states: $|0\rangle_a\otimes |0\rangle_b$ (solid line), $|0\rangle_a\otimes |1\rangle_b$ (dashed-dotted line), $|1\rangle_a\otimes |0\rangle_b$ (dashed line) and $|1\rangle_a\otimes |1\rangle_b$ (dotted line) when $\chi_{ab}=0$. Remaining parameters are the same as for Fig.1.}
\end{figure}
\begin{figure}[h]
\centerline{\includegraphics[width=0.7 \textwidth]{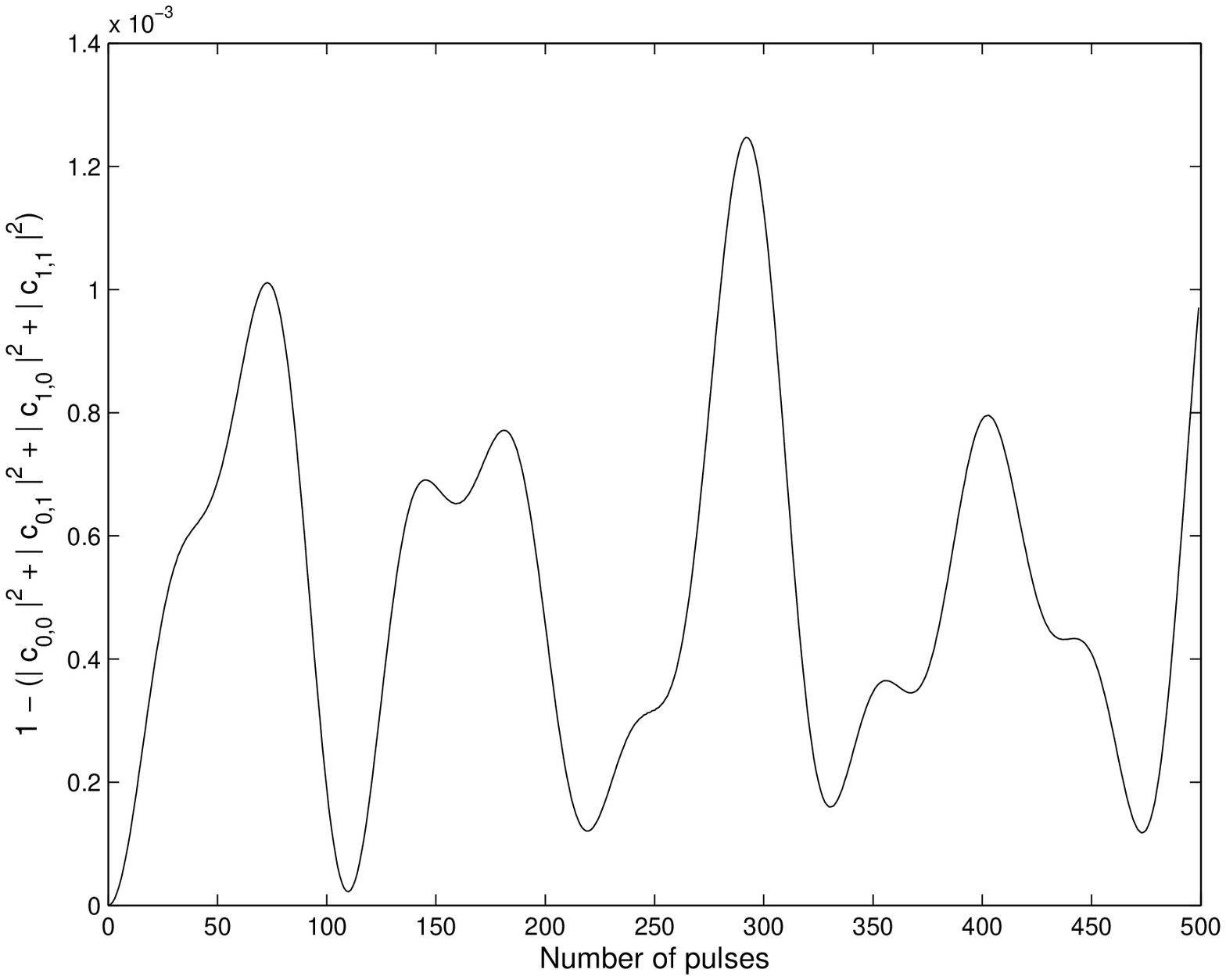}}
\caption{The same as in Fig.2 but for the probabilities from Fig.3.}
\end{figure}
Fig.3. depicts the time-evolution of probabilities for four states: $|0\rangle_a\otimes |0\rangle_b$, $|0\rangle_a\otimes |1\rangle_b$, $|1\rangle_a\otimes |0\rangle_b$ and $|1\rangle_a\otimes |1\rangle_b$, when the  cross-term is neglected \textit{i.e.} $\chi_{ab}=0$. Fact of vanishing of the last term  in Hamiltonian (\ref{eq1}) changes the situation considerably. Although our system still behaves as NQS but for this case the energies of eigenstates of $\hat{H}_{NL}$ become different from those discussed above and hence,  state $|1\rangle_a\otimes |1\rangle_b$ becomes involved in the system's evolution, as well. This state corresponds to the eigenenergy which is the same as for other three states. Analogously to the previous case, sum of probabilities corresponding to the four states considered here is close to the unity. Fig.4 shows that the deviation of this sum from the unity is sufficiently small and hence, other than the four states mentioned above can be neglected. It is seen that time-evolution of the probabilities shown in Fig.3 is more complicated than that depicted in Fig.1. Nonetheless, the system behaves as \textit{qubit-qubit} one, similarly as that with cross-coupling term. Again, only two states in each mode are involved -- vacuum and one-photon states.

It is seen from Fig.3 that  crossing of the pairs of probabilities, leading to the \textit{Bell states} generation (here, for the pairs ($|1\rangle_a\otimes |0\rangle_b$,  $|0\rangle_a\otimes |1\rangle_b$) or ($|0\rangle_a\otimes |0\rangle_b$, $|1\rangle_a\otimes |1\rangle_b$)) does not appears here. For the case when cross-coupling is neglected, such crossing effects exist only for the pairs of states: ($|0\rangle_a\otimes |0\rangle_b$,  $|1\rangle_a\otimes |0\rangle_b$) and ($|0\rangle_a\otimes |1\rangle_b$,  $|1\rangle_a\otimes |1\rangle_b$) -- it leads to the product states generation. Nonethelesss, one should keep in mind that it is potentially possible to find maximally entangled states for the values of parameters different from those considered here. However, since this paper is devoted rather to the presentation of the method of numerical simulation of the dynamics of two-mode NQS than discussing the properties of nonlinear coupler, considerations concerning entangling possibilities of such couplers will be a subject of separate article. 

\section{Summary}
We presented here a simulation method that allow to model quantum dynamics of kicked nonlinear quantum scissors' (NQS) systems. Discussed method is based on the application of the unitary evolution operators on the wave-function describing considered quantum system. Presented considerations concerns two-mode systems. We showed how to construct the both: the wave function and operators describing two-mode fields, expressed  in $n$-photon basis.  It can be easily done with use matrix orientated programming languages such as Matlab or its free clone \textit{Octave}.

In this paper we applied discussed method for finding time-evolution of the model involving two mutually interacting nonlinear quantum oscillators (Kerr nonlinear coupler) excited by a series of ultra-short external coherent pulses.  In particular, we discussed and compared two kicked nonlinear coupler models. For one of them cross-coupling was present whereas for the second model this coupling was omitted.  We showed that despite the simplicity of considered method one can obtain non-trivial results. In particular, it was shown that the quantum evolution of the system remains closed within a two-qubit Hilbert space if external excitation is weak. This result is true for the both discussed models. Moreover, for the coupler with cross-coupling two maximally entangled states (\textit{Bell states}) can be generated.  


\end{document}